\documentclass[prb,twocolumn,aps,showpacs, superscriptaddress]{revtex4}

\usepackage{graphicx}
\usepackage{bm}
\usepackage{amssymb}
\usepackage{amsmath}
\usepackage{subfigure}
\usepackage{tikz}

\begin{document}

\title{The classical $J_1$-$J_2$ Heisenberg model on the Kagome lattice}
\author{Martin Spenke}
\affiliation{Bethe Center for theoretical physics, The University of Bonn, Bonn, Germany}
\author{Siegfried Guertler}
\email{guertler@th.physik.uni-bonn.de,siegfried.guertler@gmx.net}
\affiliation{Bethe Center for theoretical physics, The University of Bonn, Bonn, Germany}
\date{\today }

\begin{abstract}
Motivated by earlier simulated annealing studies and materials with large spin on the Kagome lattice, we performed large scale parallel tempering simulations on the Kagome lattice for the
extended classical Heisenberg model including next nearest neighbor interactions. We find that even a small inclusion of a $J_2$ term induces anti-ferromagnetic order which prevails in the thermodynamic limit. The magnitude of this effect is surprising. While at $J_2=0$ the finite-size behaviour does not suggest a phase-transition, at other points the numerical result is consistent with one. Close to $J_2=0$ and for a positive sign of $J_2$ two subsequent phase-transitions/crossovers are found, one of them connecting to the crossover for the $J_2=0$ case, shedding light to the pure case. The universality classes of the transitions were explored.
\end{abstract}

\pacs{75.10.Hk,75.10.Jm,75.50.Ee}
\maketitle

\section{Introduction} 

The 2D Kagome lattice being the underlying lattice of a number of materials, is of great interest due to its highly frustrated geometry with respect to the antiferromagnet. A number of these materials have larger spin. E.g. SrCr$_{8-x}$Ga$_{4+x}$O$_{19}$ \cite{RAM1} with spin-$\frac{3}{2}$ or KFe$_3$(OH)$_6$(SO$_4$)$_2$ \cite{NISH1,MATAN} with spin-$\frac{5}{2}$. With growing spin we expect the model to exhibit some behaviour captured in the classical model. This model has been studied by Chalker et al \cite{CHAL1} by Monte-Carlo, spin-wave approximation and low temperature expansion. Here the ``order-by-disorder'' scenario with coplanar ordering was proposed.\cite{VIL} 
Reimers et al \cite{REIM} have performed Monte-Carlo runs more systematically and could confirm some of the predictions, particularly coplanar ordering. Chandra et al and Ritchey et al have extended the work and studied anisotropy.\cite{CHAN1,RIT1} A cross-over for the pure case and a second-order phase transition for any inclusion of anisotropy was proposed. More recently the classical Heisenberg model has been investigated by Zhitomirsky \cite{ZHIT1} with a simulated annealing method. He found a strange finite-size behaviour in the specific heat, inconsistent with a phase-transition. The antiferromagnetic order was shown to vanish in the thermodynamic limit and the so called ``octopular-order'' parameter, a third rank tensor, was proposed to describe the low temperature phase. Subsequently Gvozdikova et al extended the studies by inclusion of a magnetic field in the Hamiltonian \cite{GVOZ} following earlier work of Zhitomirsky.\cite{ZHIT2} Another relevant study is by Korshunov, who studied the XY-$J_1-J_2$ model on the Kagome and found magnetically and topologically ordered phases.\cite{KOR}\\

Classical models with a continuous symmetry are generally not expected to show any long-range order at finite temperature, except for topological order, due to the theorem of Mermin and Wagner.\cite{WAG} For frustrated models the situation is more subtle, as a discrete symmetry can be induced and this symmetry can in principle be broken. This possibility was first suggested by P. Chandra et al\cite{CHANIS}, and has been observed numerically by Weber et al,\cite{WEBER}. Note that this does not violate the Mermin-Wagner theorem, as the symmetry being broken is not a continous one, but an additional discrete one. Meanwhile a mathematical proof is published about such type of models,\cite{BIS} showing that such induced symmetry can be broken. In the classical nearest neighbor Heisenberg-model on the Kagome (CKLNNHM), all studies of the past are not consistent with a phase-transition at finite temperature. Features in the observables, such as a peculiar finite size behaviour in the specific heat have been observed in Monte-Carlo simulations.\cite{ZHIT1} We refer to this point as ``crossover-temperature'' in our manuscript.\\

Here we investigate the extended Heisenberg model by inclusion of a next-nearest neighbor (NNN) coupling term. The motivation for this is three-fold: First, in real materials a small NNN coupling is expected; second, from observing the trends towards $J_2 \rightarrow 0$ new insight to the pure model might be possible; and third, this model couples to three underlying sub-lattices which are again of Kagome type (see Fig. \ref{KNNN}). This 
is therefore the first study of frustration which considers the coupling of two frustrated lattices which to some extent ``unfrustrate'' them as the inclusion of the $J_2$ lifts some of the degeneracies.
Our results show that even a very small $J_2$-term enables an anti-ferromagnetic order. This order is of $q=(0,0)$ type for AF-coupling of $J_2$, and of $\sqrt{3} \times \sqrt{3}$ type in the other case. For the pure case ($J_2=0$) the crossover temperature is tiny ($T_k\approx0.003$) this changes quite dramatically upon switching on the $J_2$-term (e.g. we obtain $T_c=0.045\pm0.001$ for $J_2=0.02$). Away from $J_2=0$ the finite-size scaling is consistent with a phase-transition. The universality class could be classified to some extent and was found to be of the Ising type for $J_2\approx\pm0.02$. For positive $J_2$ which is close to $J_2=0$ two peaks in the specific heat are found numerically, suggesting two phase-transitions. Studying the limiting behaviour from the other side towards $J_2=0$ appears to be numerically harder. Here the phase-transition is consistent with a first-order transition, which is apparently the reason for the numerical difficulty.

\section{Model} 

\begin{figure}
     \centering
    \includegraphics[width=0.95\columnwidth]{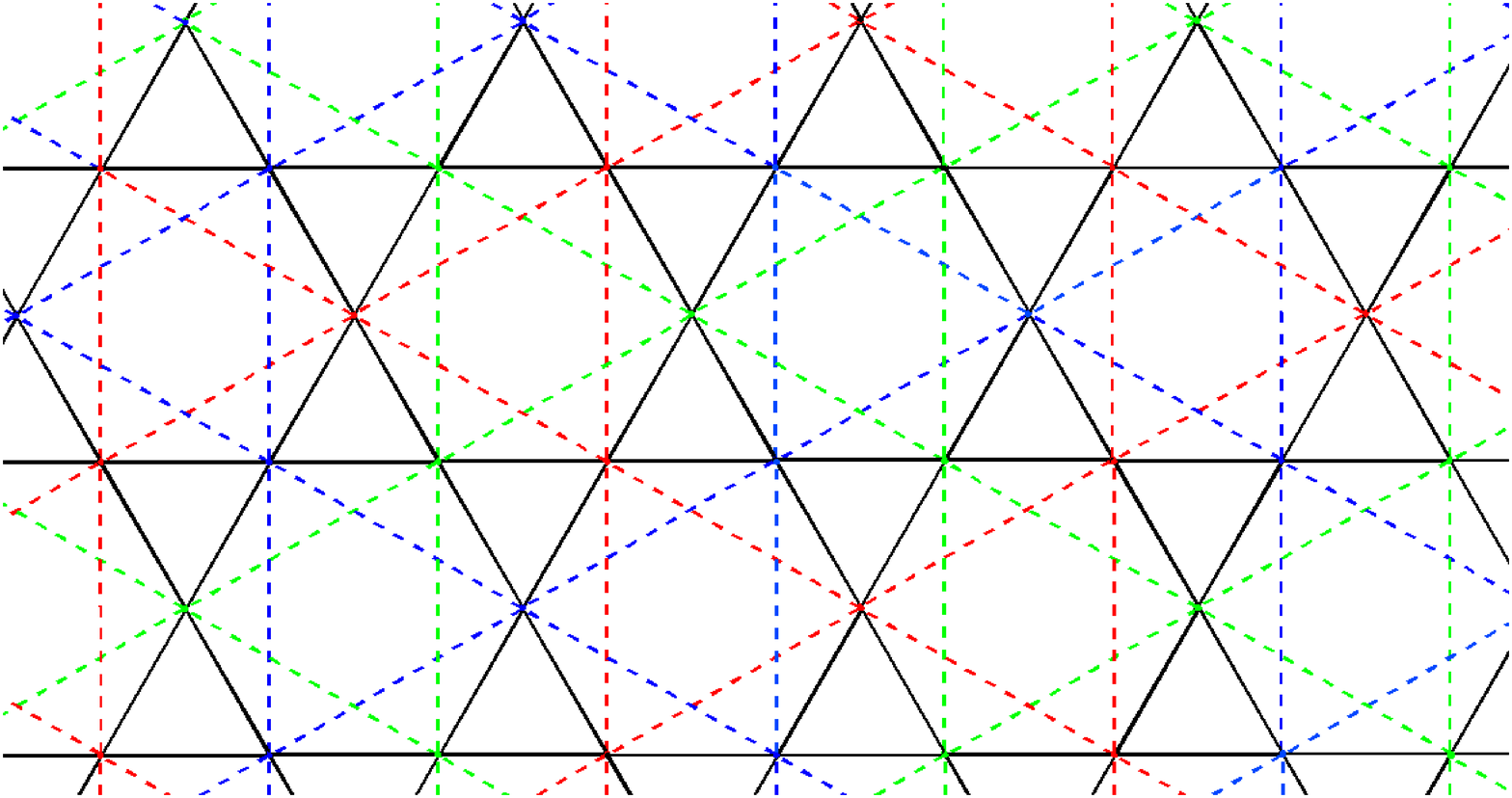}
    \caption{(Color online) The Kagome lattice, including the next nearest neighbor bonds (in dashed lines). The three underlying NNN-Kagome sub-lattices are shown in different colours.}
    \label{KNNN}
\end{figure}

\begin{figure}
  \begin{center}
    \begin{tabular}{cc}
      \resizebox{40mm}{!}{\includegraphics{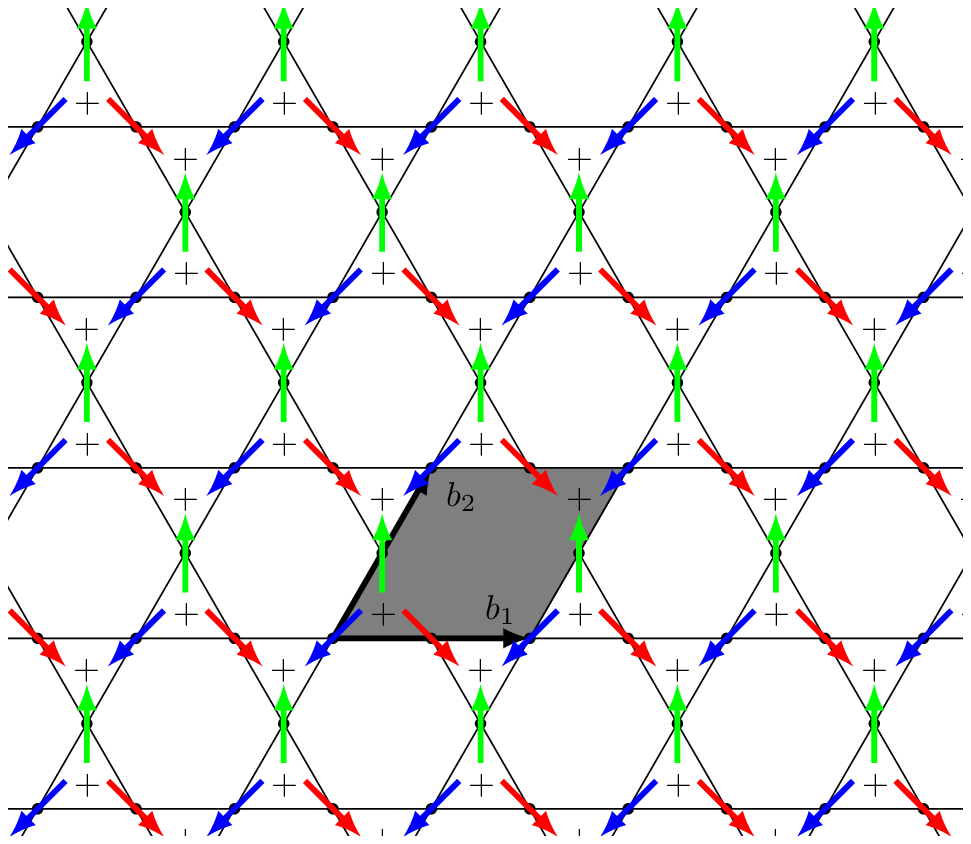}}  &
      \resizebox{40mm}{!}{\includegraphics{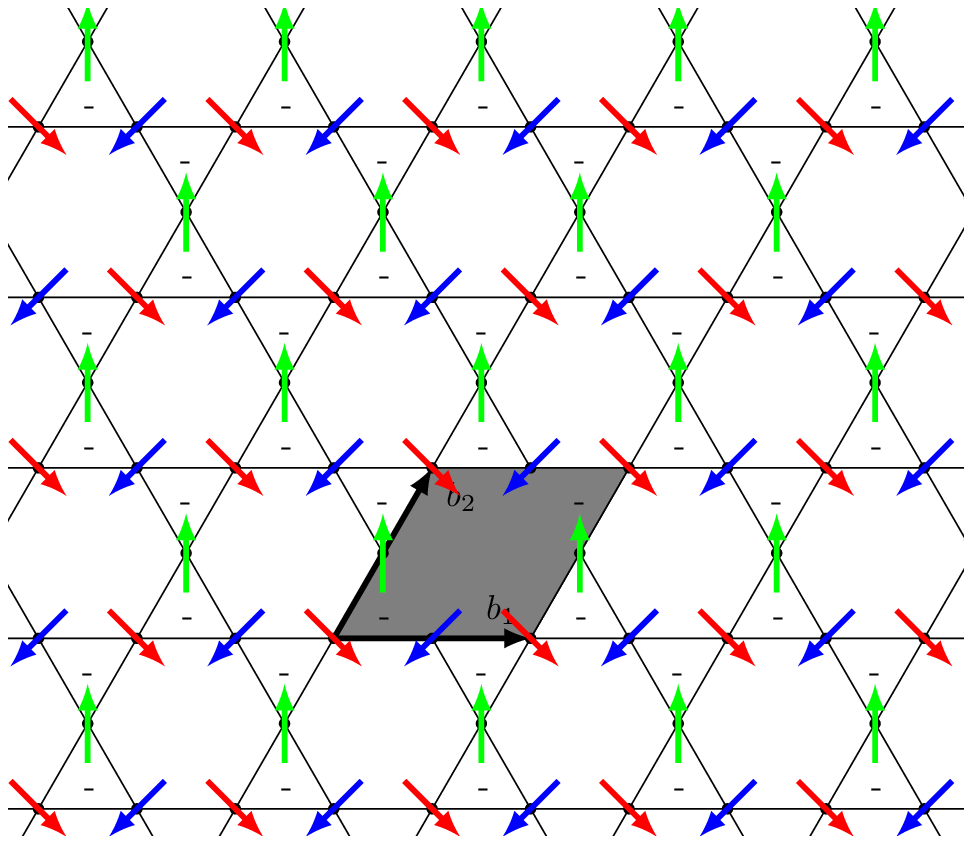}}  \\
      \resizebox{40mm}{!}{\includegraphics{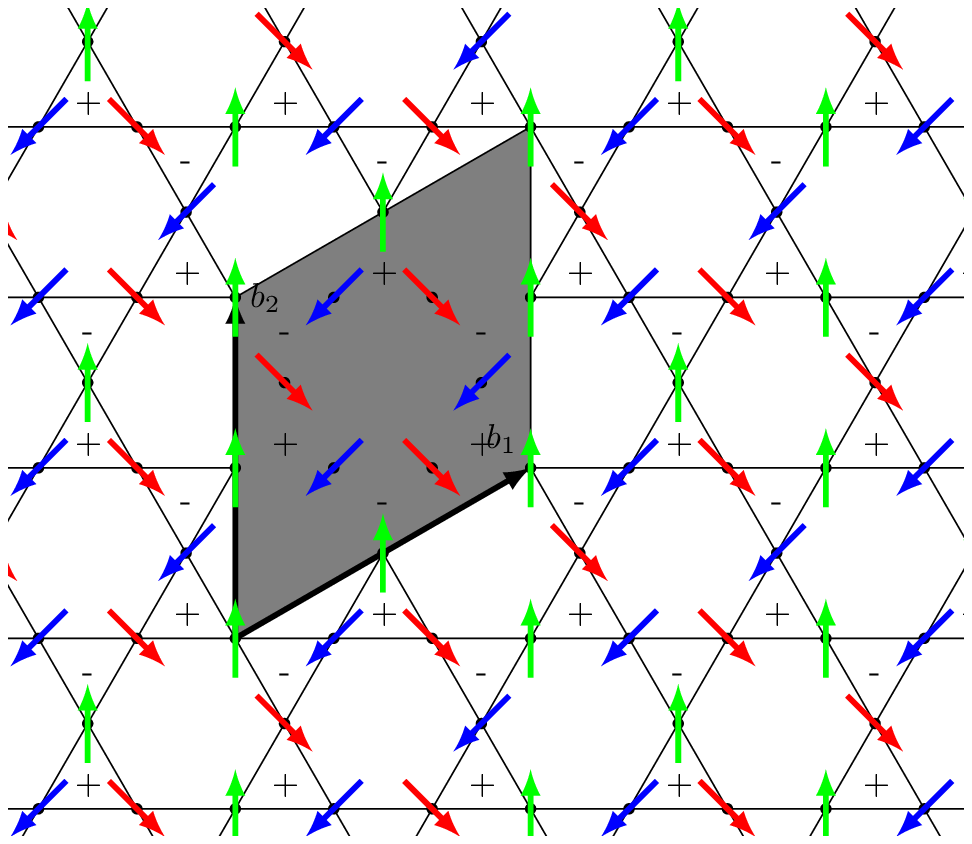}}  &
      \resizebox{40mm}{!}{\includegraphics{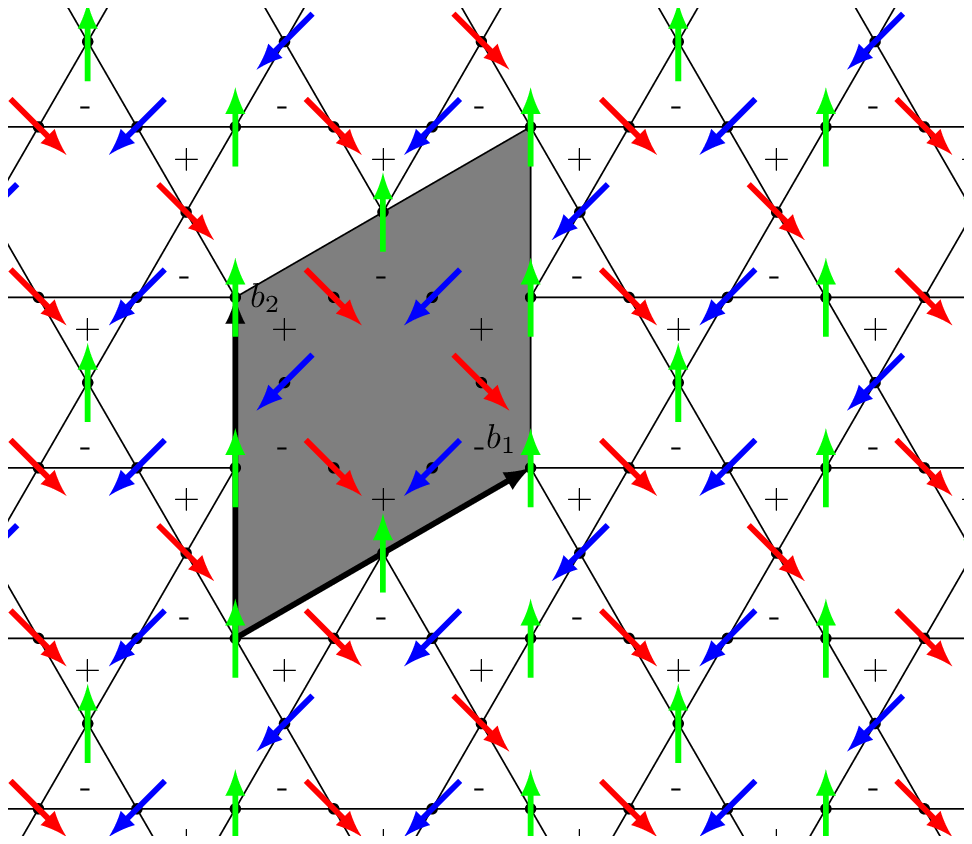}} 
    \end{tabular}
    \caption{(Color online) Possible anti-ferromagnetic (AFM) sub-lattice ordering pattens. Top: AFM ordering with $q=(0,0)$ with an unit cell similar to the one of the Kagome lattice, in its two possible forms with positive (top left) and negative (top right) chiralities. We refer this state as AFM I in the text. Below: AFM ordering with $q=(4 \pi/3,0)$. The corresponding enlarged unit cell is indicated in grey. This AFM state possess two possible forms  as indicated in the two pictures. We refer to this order as AFM II from now on.}
    \label{AFM12}
  \end{center}
\end{figure}

\begin{figure}
     \centering
    \includegraphics[width=0.8\columnwidth]{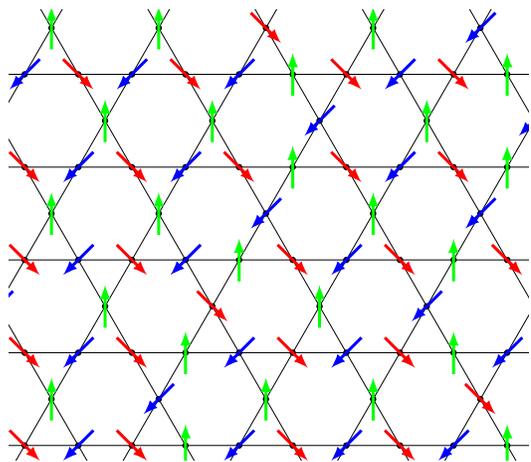}
    \caption{(Color online) An octupolar ordered state.}
    \label{OCC}
\end{figure}

From earlier work by the authors mentioned in the introduction, it was shown that the CKLNNHM is very hard for a numerical investigation: The interesting temperature region is very low, and the low temperature phase is very difficult to bring into equilibrium.
The extension of further neighbor interaction lift some degeneracies and therefore taking away some of the numerical ``hardness''. 
Our model Hamiltonian reads:

\begin{equation}
H=-J_1\sum_{\langle ij\rangle}\mathbf{S}_i\cdot\mathbf{S}_j+J_2\sum_{( ij)}\mathbf{S}_i\cdot\mathbf{S}_j
\end{equation}

where the $J_1$ term couples between nearest neighbor pairs (indicated by solid lines in Fig. \ref{KNNN}) and the $J_2$ term between next-nearest neighbor pairs (indicated by dashed lines in Fig. \ref{KNNN}). In the following we give $J_2$ in units of $J_1$. $\mathbf{S}$ is a classical spin operator. When $J_2=0$ we recover the nearest neighbor model. In this case there is a large degeneracy of coplanar and non-planar states. Let us discuss the possible ground states in this case: Ground states can be found by rewriting the Hamiltonian from a nearest neighbor sum to a sum over all triangles $\alpha$ as:

\begin{equation}
H=-J_1\sum_{\langle ij\rangle}\mathbf{S}_i\cdot\mathbf{S}_j =\frac{J_1}{2}\sum_{\alpha}(\mathbf{S}_1+\mathbf{S}_2+\mathbf{S}_3)^2+c
\end{equation}

where $c$ is a constant term. From this sum one realizes the condition to minimize the energy: If the sum of each individual triangle is zero, this is a minimum energy state. There are two groups of such states: the coplanar ones, where all 3 spins lie in one plane, and the non-coplanar ones, where this is not the case. We will first discuss the coplanar states and briefly mention how to obtain the non-planar ones from them. The coplanar states contain two types of antiferromagnetic sub-lattice order pattens (see Fig. \ref{AFM12} thereafter AFM-states). There is a continuous $O(2)$ symmetry for these states as a simultaneous rotation of all spins in the plane keeps the energy constant. Both variants of AFM type states are two-fold degenerate: For the $q=(0,0)$ state (from now on ``AFM I'' state) there is one version with only positive chiralities and one state with only negative chiralities. Chirality refers to the clockwise or anti-clockwise type of rotation of spins when circumferencing an elementary triangular plaquette. For the other ($q=(4 \pi/3,0)$ from now on dubbed as ``AFM II'' state - this state is usually referred to as the $\sqrt 3 \times \sqrt 3$ structure) AFM state there is one version with all the up-triangles having positive chirality and all the down-triangles having negative chirality and one version where the opposite is the case. For an illustration please see Fig. \ref{AFM12}. 
The above condition for minimizing the energy is not only full-filled by the two AFM states, but by all states having 3 spins in 
one plane arranged among the plaquette with 120 degrees. Starting with one sub-lattice, the angles are fixed, but the arrangement of the 3 spins 
can be varied. For an arbitrary arrangement fullfilling the 120 degrees condition there is no regular patten in chiralities. One may label the 3 spins with 3 colors e.g. red, blue and green, and compose pattens where one coloured site has only nearest neighbors of another colour. All coplanar states can be constructed in such a manner. See Fig. \ref{OCC} for illustration of a general planar state.\\

Taking a coplanar states as a starting point, one may use deformations leaving the energy unchanged to construct non-planar states: If e.g. we consider the two AFM states of Fig. \ref{AFM12} there are two types of defects: The weather-vane defect for AFM II, here one rotates all inner spins of a hexagon along the common axis given by the outer spins, and a line-defect for the AFM II, where one can rotate a whole line, by the axis given by the neighbouring spins. The energy is kept constant in both cases. Considering an arbirary situation (a coloured state which is neighter AFM I nor AFM II), one can do such operations among two coloured lines. From the above discussion one realizes that the ground state is highly degenerate and disordered. As mentioned in the introduction, previous studies proposed an order-by-disorder scenario: For $T\ne0$ entropic selection favours coplanar states over the other states. As an order parameter originally a nematic order parameter was considered in the earlier studies. This measures the alignment of spins with respect to a common plane. The general patten of 120 degree alignment, discussed above and being an additional symmetry is not measured by such an order parameter. The octopular tensor was proposed as a better choice to measure states as the one in Fig. \ref{OCC}. We follow this approach.\\

Finally we turn to the situation with $J_2\ne0$: In this case we bias the lowest energy state depending on the sign chosen for $J_2$ towards the two AFM states (which are still degenerate with respect to rotation and have an additional two-fold degeneracy as discussed above). Therefore we lift most of the degeneracies. When being in one perfectly ordered AFM state of one manifestation of the chirality, one needs to overcome an energy barrier to reach the other one. This is different from the situation at $J_2=0$, as there the two states can be reached with formation of domain walls costing zero energy. Therefore at $J_2\ne0$ there is a discrete $Z_2$ symmetry for planar states.
The building blocks of up and down triangles possessing '+' and '-' chiralities can therefore be understood of taking the role of Ising variables.
Note that for planar states the situation is the same as in the XY-model on the Kagome lattice being discussed in detail in Ref. \onlinecite{KOR}.

\section{Method} 

We employ an improved parallel tempering (PT) approach,\cite{SW1,PTR,KTT} based on the ALPS libraries and parapack.\cite{ALPS} We have tested several optimizations from the literature and finally used the following method: A suitable set of temperatures is chosen, then auto-correlation-times are measured to optimize the frequency of exchange steps between neighbouring replicas. This is the so called ``$\tau$-method'' of Bittner et al.\cite{BIT1} As updates we used restricted updates and perform over-relaxation.\cite{ZHIT1} This method shows roughly 100 times faster convergence compared with canonical Monte Carlo and naive updates. Typically we used around the order of $10^5$ sweeps for thermalization and another $10^7$ sweeps for measurements. One PT simulation consisted of 90 to 180 replicas. The exact amount of sweeps for thermalization and measurements and the number of replica varies largely depending on the regime of interest. The AF-coupled region close to $J_2=0$ is very hard. As discussed below we believe this is a consequence of the nature of the phase-transition at this point. We have tested our program intensively by comparing our data against earlier published studies.\cite{ZHIT1,DIEP,ISK} The $J_2$ term was tested by noting the structure of the NNN-lattice being of 3 independent Kagome lattices of smaller size (see Fig \ref{KNNN}). Choosing the couplings $J_1=0$, $J_2=1$ should therefore give similar curves in the measurements compared to choosing $J_1=1$, $J_2=0$ when considering the change in the volume-size $N$. Our tests left no doubt our program works correctly.\\

To identify possible phase-transition points, we measure the specific heat $c_v$. This is an quantity which should show a feature at a transition or even a crossover point, independently of the kind of order-parameter:

\begin {equation} 
c_v = \frac {\langle E^2\rangle-\langle E \rangle^2}{NT^2}  
\end{equation}

Following earlier literature we measure the octupolar order parameter $T^2_{OP}$ as mentioned above and investigated by Zhitomrisky:\cite{ZHIT1,RIT1}

\begin{equation}
\bigl(T^{\alpha\beta\gamma}\bigr)^2 = \frac{1}{N^2} \sum_{i,j} 
\Bigl[ \langle ({\bf S}_i\cdot{\bf S}_j)^3\rangle  - 
\frac{3}{5}\langle {\bf S}_i\cdot{\bf S}_j\rangle  \Bigr].
\end{equation}

with the limiting value $\frac{1}{4}$ in the limit $T\rightarrow 0$. See Fig. \ref{OCC} for an example of an octupolar ordered state. In addition we measure antiferromagnetic long-range order defined as:

\begin{equation}
m_{\rm AF}^2 = \frac{6}{N^2} \sum_{l,i,j} 
\langle {\bf S}_{li}\cdot {\bf S}_{lj} \rangle 
e^{i{\bf q}({\bf R}_i-{\bf R}_j)} 
\label{mAF}
\end{equation} 

where the first index $l$ labels the site within the unit-cell, the other indexes label the xy-position of the unit cell within the lattice and 
$R_i$ is the position of the unit-cell within the underlying triangular lattice. For $q$ we choose the two vectors corresponding to the ordering 
pattens for the AFM I and AFM II, $q=(0,0)$ and $q=(4 \pi/3,0)$. Among the two orders the AFM I order has been shown to be very difficult to equilibriate for the pure case, particular cooling and warming runs starting from this state differ significantly.\cite{REIM} On the other hand there have been speculation about possible AFM II order or correlations in the pure model,\cite{CHAL1,HUSE} which have been shown to vanish for any non-zero temperature performed numerically.\cite{ZHIT1}
Another quantity of possible interest is the spin-stiffness:

\begin{equation}
\rho_s = -\frac{\sqrt{3}}{2N}\Bigl\{ \frac{1}{3}\langle E\rangle+
\frac{J^2}{T} \Bigl\langle \Bigl[ \sum_{\langle ij\rangle}
({\bf S}_i\times{\bf S}_j) \cdot \hat{\bf e}_{\alpha}
(\hat{\bf e}_{\beta}\cdot \hat{\bf \epsilon_{ij}}) \Bigr]^2 \Bigr\rangle
\Bigr\}
\end{equation}

where $\hat{ \vec e}_{\alpha}$ and $\hat {\vec e}_{\beta}$ are both
arbitrary unit vectors.
$\hat{\bf \epsilon_{ij}}$ is the unit vector in the direction of the
bond between $i$ and $j$. To investigate possible further structure within the spin-ordering, we measured the spin-structure factor $S({\bf q})$ for a few particular values defined as:

\begin{equation}
S({\bf q}) = \frac{1}{N} \sum_{i,j} 
\langle {\bf S}_{i}\cdot {\bf S}_{j} \rangle 
e^{i{\bf q}({\bf r}_i-{\bf r}_j)} 
\label{Sq}
\end{equation}

where $r_i$ labels the real space position, and ${\bf q}$ refers to the momenta.

\section{Results} 

\begin{figure}
     \centering
    \includegraphics[width=0.95\columnwidth]{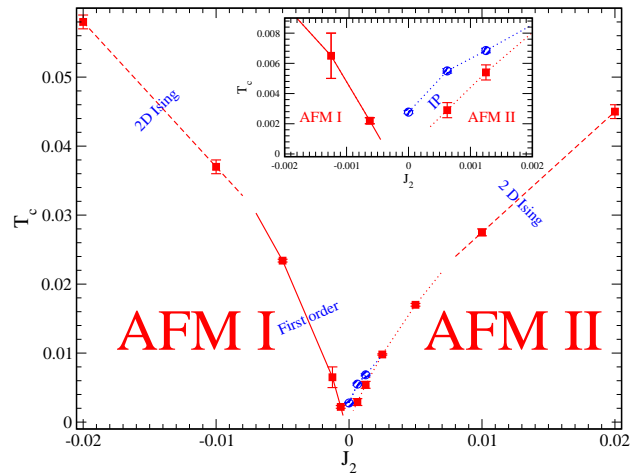}
    \caption{(Color online) Phase-diagram of the extended Heisenberg-model on the Kagome lattice: Shown are the estimated transition (for non-zero $J_2$) and  crossover (for $J_2=0$) points corresponding to antiferromagnetic and octopular order. AFM I labels the phase with the $q=(0,0)$ antiferromagnetic order, AFM II labels the phase with the $\sqrt 3 \times \sqrt 3$ structure - in these two phases the octupolar order is found in addition to the antiferromagnetic one, IP labels the region between the two peaks of the specific heat for the positive $J_2$-branch close to $J_2=0$. The solid line indicates the transition line consistent with first order scaling laws. The dashed lines are those transition lines consistent with the 2D Ising universality class. For the dotted line the classification failed.}
    \label{PHASE}
\end{figure}

Our main result is summed up in the form of a phase diagram in Fig. \ref{PHASE}. As established below at any $J_2 \ne 0$ we see a finite-size behaviour consistent with a finite temperature phase transition. Starting from $T_k$, the crossover temperature at $J_2=0$, we see that a small value of $J_2$ does not only enable a phase transition but has a huge impact on the point of transition. As expected depending on the sign of $J_2$ we achieve ordering of the AFM I or AFM II type. The transition has been found to be of second order with 2D Ising universality class for $|J_2|>0.0075$ and of first-order for $0>J_2>=-0.0075$. For tiny but positive $J_2$ two peaks in the specific heat are observed and a classification in terms of universality failed.\\

\begin{figure}
     \centering
    \includegraphics[width=0.95\columnwidth]{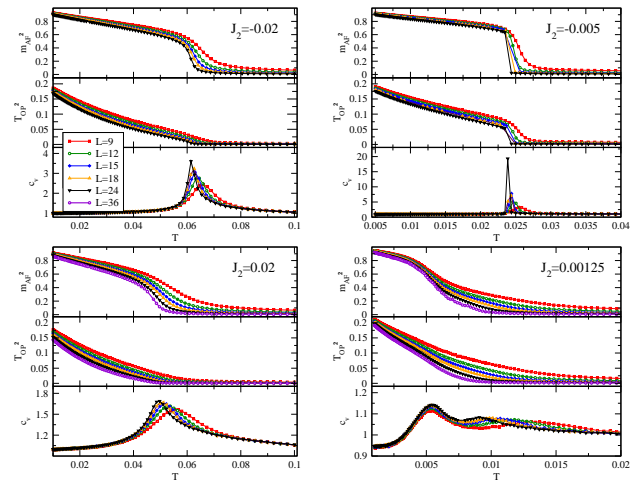}
    \caption{(Color online) The finite-size behaviour of the most important measurements for four values of $J_2$. Error-bars are smaller than the symbol-size.}
    \label{OBS4J2}
\end{figure}

\begin{figure}
     \centering
    \includegraphics[width=0.83\columnwidth]{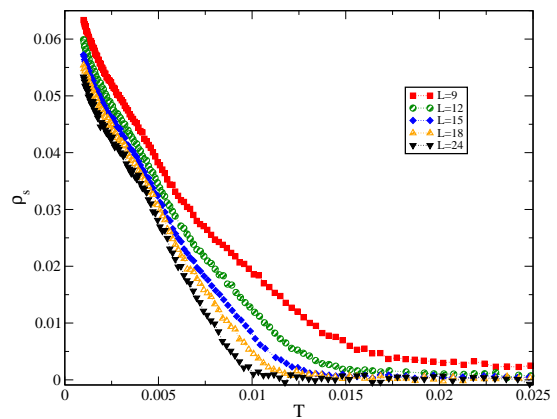}
    \caption{(Color online) Spin-Stiffness for $J_2=0.00125$ and several system sizes. Error-bars are smaller than the symbol-size.}
    \label{STIFF}
\end{figure}

The finite-size behaviour is summed up in Fig. \ref{OBS4J2}. Here we show the three most relevant observables $T^2_{OP}$, $m_{AF}$ and $c_v$ for four values of $J_2$. These are snapshots for the interesting regions in the phase-diagram. For values of $|J_2| > 0.0075$ (in the picture we choose $J_2=\pm 0.02$) there is a phase transition between the disordered and AFM phase. The specific heat develops a peak which gets narrower and higher with increasing lattice size. The AFM phase is of AFM I type for negative $J_2$ and AFM II type for positive $J_2$. The position of the peak appears where the measured AFM-ordering has its inflection point. At the same position $T^2_{OP}$ appears to have its onset. For smaller values of $J_2$ the nature of the transition changes: For e.g. $J_2=-0.005$ the peak of the specific heat grows much faster compared to the $|J_2| > 0.0075$ cases. The order-parameters $m_{AF}$ and $T^2_{OP}$ have a more sudden change between the two phases. In this part there is very slow convergence when simulating. To us it appears to be a first-order transition which we elaborate below. 
In the region close to $J_2=0$ but with positive sign for $J_2$, we find another interesting feature: For e.g. $J_2=0.00125$ (see Fig. \ref{OBS4J2}) we find two peaks in the specific heat, the taller one matching to the inflection point of $m_{AF}$ the second, less dominant one matching to the onset of the $T^2_{OP}$ and $m_{AF}$ order. When observing the finite-size effect, we notice that the bigger peak stays almost stationary at the same position with respect to the temperature, when increasing the system-size, while the smaller peak moves towards smaller temperature with increasing system-size. An extrapolation suggests the two peaks to stay separated in the thermodynamic limit. Investigating the region between the two peaks we found that both $T^2_{OP}$ and $m_{AF}$ orders are reduced in the thermodynamic limit: an extrapolation suggests here the $T^2_{OP}$ order clearly to vanish, while for the AFM a small value may remain for infinite lattice sizes. This ``phase'' therefore seems to resemble some of the behaviour compared to the $J_2=0$ case, which shows similar features in $T^2_{OP}$ and $m_{AF}$. Two differences of this intermediate ``phase'' compared to the situation at $J_2=0$ can be seen: First, in the $J_2=0$ case the $m_{AF}$ order is vanishing faster than the $T^2_{OP}$, here this trend is opposite. Second, the finte size trend for the upper specific heat (the smaller one) seems still consistent with one at a phase-transition as this peak still seem to grow taller and narrower with larger system-size. Since the two peaks are quite close to each other, it is hard to be clear about the trend of this second smaller peak, resting on top of the slope of the bigger peak. This upper transition-line separates at around $J_2=0.002$ from the main-phase boundary and continues to the point of crossover for the $J_2=0$ case. Using the transistion points to extrapolate to the regime where simulation is no more possible it one notices:
The lower transition line for the positive $J_2$ branch between the AFM 2 state and the disordered phase tend to $T\rightarrow0$ for $J_2\rightarrow0$, while in the other branch the AFM 1 order seems to vanish before reaching the $J_2=0$ point. This is consistent with earlier claims of a $T=0$ critical point for the AFM 2 transition.\cite{HUSE,REIM} For $J_2=0.00125$ we checked to which of the two peaks the spin-stiffness matches (for the other cases, the spin-stiffness generally speaking has the same behaviour in terms of onset and infliction point compared to the $m_{AF}$ and $T^2_{OP}$). This is shown in Fig. \ref{STIFF}. We observe again a similar trend compared with the $T^2_{OP}$ and $m_{AF}$ measurements: The onset matches the first peak, while the infliction point is matching the broad peak.\\

\begin{figure}
     \centering
    \includegraphics[width=0.95\columnwidth]{L24NEGV5.eps}
    \caption{(Color online) Measurements for the negative $J_2$ case of the antiferromagnetic order ($m_{AF}$), the octopular order ($T^2_{OP}$) and the specific heat $c_v$ for several values of $J_2$.
The system size was $L=24$. Error-bars are indicated if they exeed the symbol size.}
    \label{NEG24}
\end{figure}

\begin{figure}
     \centering
    \includegraphics[width=0.95\columnwidth]{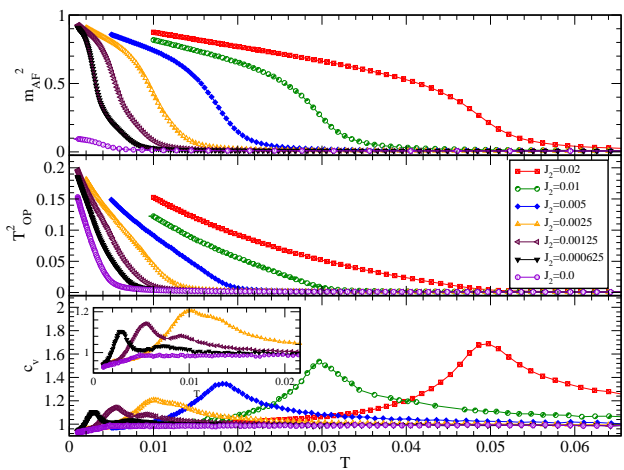}
    \caption{(Color online) Measurements for the positive $J_2$ case of the antiferromagnetic order ($m_{AF}$) the octopular order ($T^2_{OP}$) and the specific heat $c_v$ for several values of $J_2$.
The system size was $L=24$. Error-bars are smaller than the symbol size.}
    \label{POS24}
\end{figure}

\begin{figure}
     \centering
    \includegraphics[width=0.78\columnwidth,angle=270]{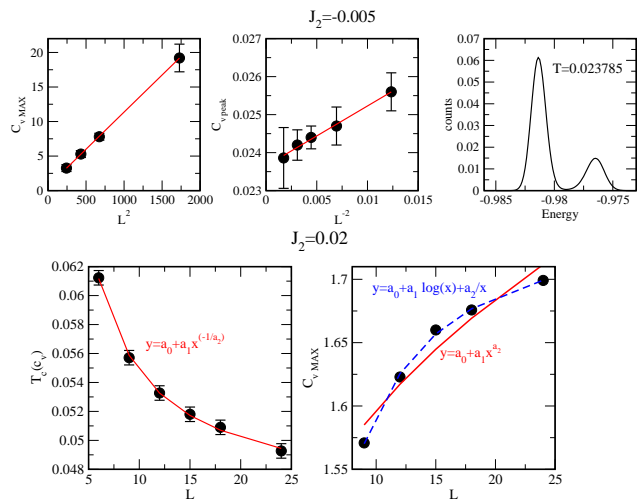}
    \caption{(Color online) Scaling fits for two points $J_2=-0.005$ (above) and $J_2=0.02$ (below). Above left: The specific heat height is found to be proportional $L^2$; Above middle: $T_c(L)$ scales with $L^{-2}$; Above right: Double peak structure in the energy at the estimated transition point: We used around 20 mill. sweeps for measurements and the system size was $L=24$; Below left: $T_c(L)$ can be fitted towards power law. Below right: Peak height shows logarithmic dependence.}
    \label{SCAL}
\end{figure}

In Fig. \ref{NEG24} and Fig. \ref{POS24} we plotted the development of the measured quantities for a larger system size ($L=24$) as a function of $J_2$. For the negative branch, one can see clearly that the nature of the transition is becoming more consistent with a first-order phase-transition when coming closer to the point $J_2=0$. In the positive branch, the specific heat is changing from a very broad single peak at large $J_2$ values to a smaller double peak close to $J_2=0$. The octupolar order $T^2_{OP}$ has a more sudden rise when tuning $J_2$ towards smaller values. At $J_2=0$ the peak in the specific heat vanishes when increasing the lattice-size, as observed by Zhitomirsky, suggesting no phase transition at this point. The AFM order vanishes at this point in the thermodynamic limit. Elsewhere, as mentioned above we note that the trend of the specific heat, seems to be consistent with a phase-transition, this is true also for the second smaller peak for positive and small $J_2$. In We tried to fit the data to scaling laws:
The points easiest to fit, are those with large values for $|J_2|$ as here the temperature for the point of transition is higher. Fitting a relation for the dependence of the estimate of $T_c$ on $L$ (we tried several ways, one of them is the peak in $c_v$) to a power law gives a value for the critical exponent of $\nu=0.99\pm 0.05$ at $J_2=0.02$ where the error-bar is estimated from different ways to obtain the $T_c(L)$. This would be consistent with a 2D Ising universality class. To test this theory we fitted the height of the specific heat. For the 2D Ising universality class such a fit should not be possible, as the divergence is logarithmic. Any power-law fit gives indeed systematic errors compared to the data, while a logarithmic fit agrees well (see Fig. \ref{SCAL} lower panel). For a value of $J_2=0.01$ we obtain $\nu=1.08\pm0.05$, and again a logarithmic divergence for the height of the specific heat. This still seem to somewhat fit in this universality class. For smaller values of the positive signed $J_2$, our data does not allow to get a consistent value for this critical exponent. Here the situation is more complicated since two peaks in the specific heat develop. As these two peaks are close to each other, it is not too surprising that finite size scaling fails. At the negative $J_2$ branch, the large $|J_2|$ value is again consistent with the Ising universality class with a value of $\nu=0.99\pm0.05$. For smaller values of $|J_2|$ of the same sign,
the transition point appears to become sharper and the specific heat diverges faster, as one would expect in a first-order transition. For a first order transition the maximum of the specific heat is expected to be proportional to the volume $L^2$ and $T_c(L)$ should scale with $L^{-2}$.\cite{FIRST} Plotting $c_{v,max}$ versus $L^2$ for the case $J_2=-0.005$ gives a linear dependence and the scaling of $T_c(L)$ agrees well with the expected $L^{-2}$ relation (see Fig. \ref{SCAL}). Therefore we see that the maximum of the specific heat is proportional to the volume and the position of the same peak is proportional to $L^{-2}$ as expected for a first order transition. We recorded energy-histograms and found the double-peak structure a typical feature of first-order transistions. Obtaining this histogram was somewhat challenging, since it converges very slowly and seem to appear only very close to the transition point (as we show in Fig. \ref{SCAL}). We needed 6 digits for the temperature to find two peaks and 20 mill. sweeps for measurements. From our observations we estimate the change from second-order to first-order behaviour taking place at around $J_2\approx-0.0075\pm0.002$, here a tricritical point is expected. There are several possible scenarios how a second-order phase transition changes to become a first-order one when tuning parameters,\cite{BINDER} an analysis of the underlying mechanism causing this, we leave for future studies.\\

\begin{figure*}
  \begin{center}
    \begin{tabular}{ccc}
      \resizebox{50mm}{!}{\includegraphics{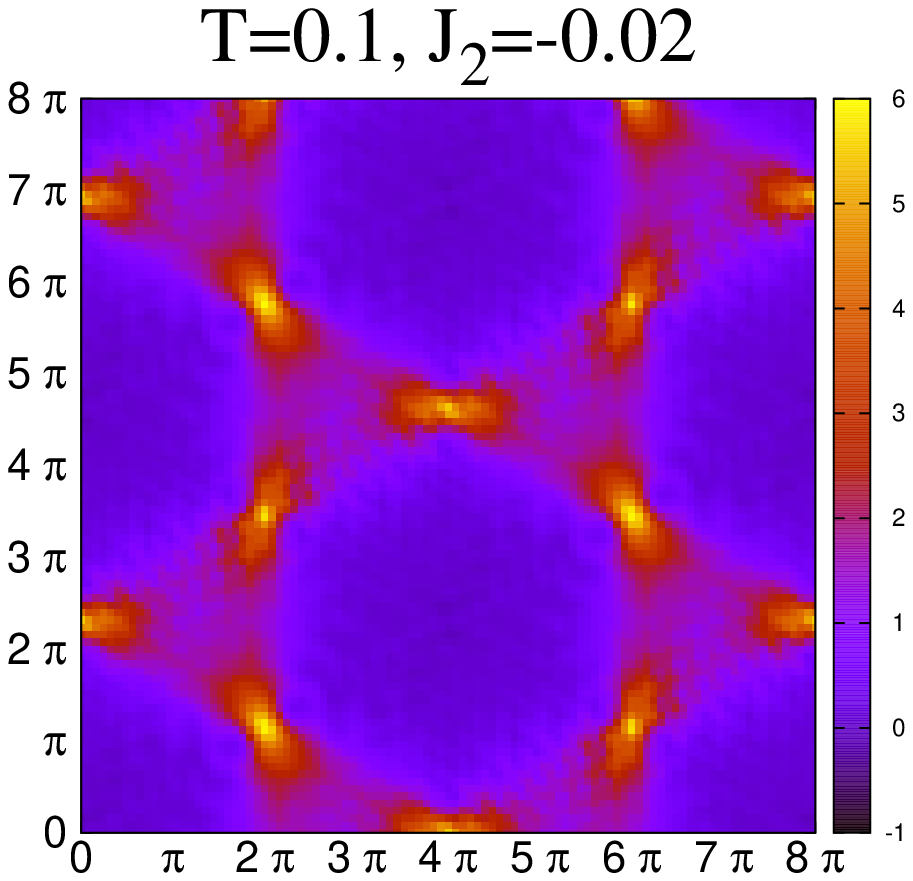}}  &
      \resizebox{50mm}{!}{\includegraphics{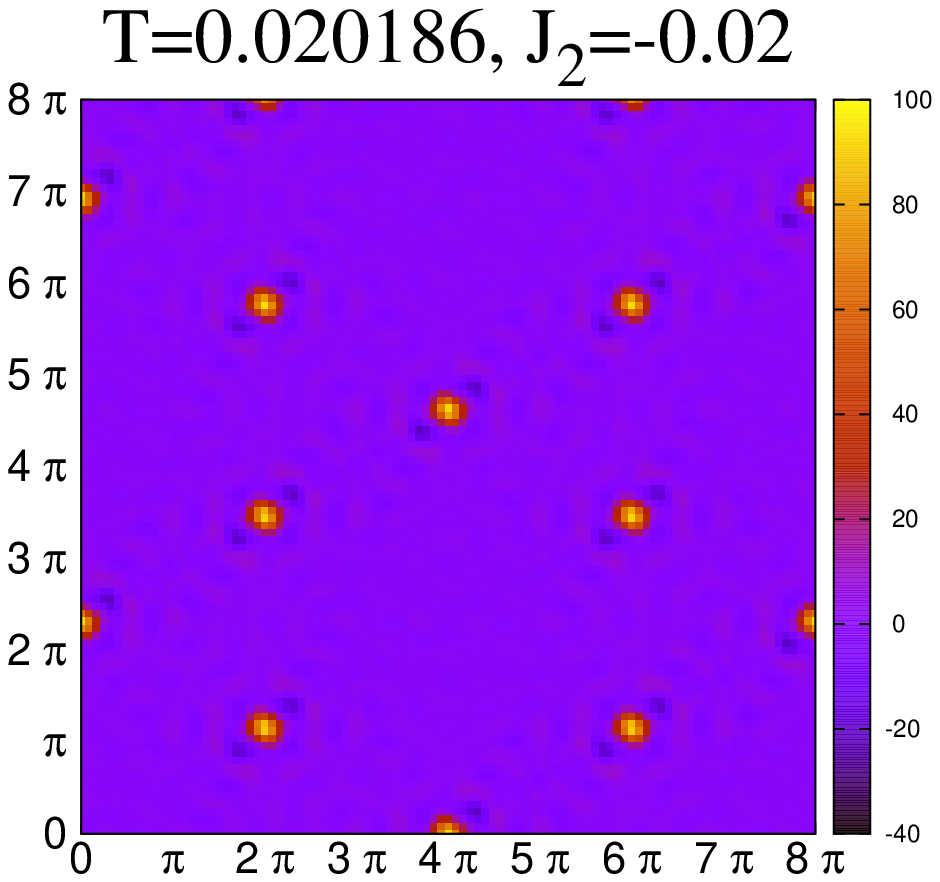}}  \\
      \resizebox{50mm}{!}{\includegraphics{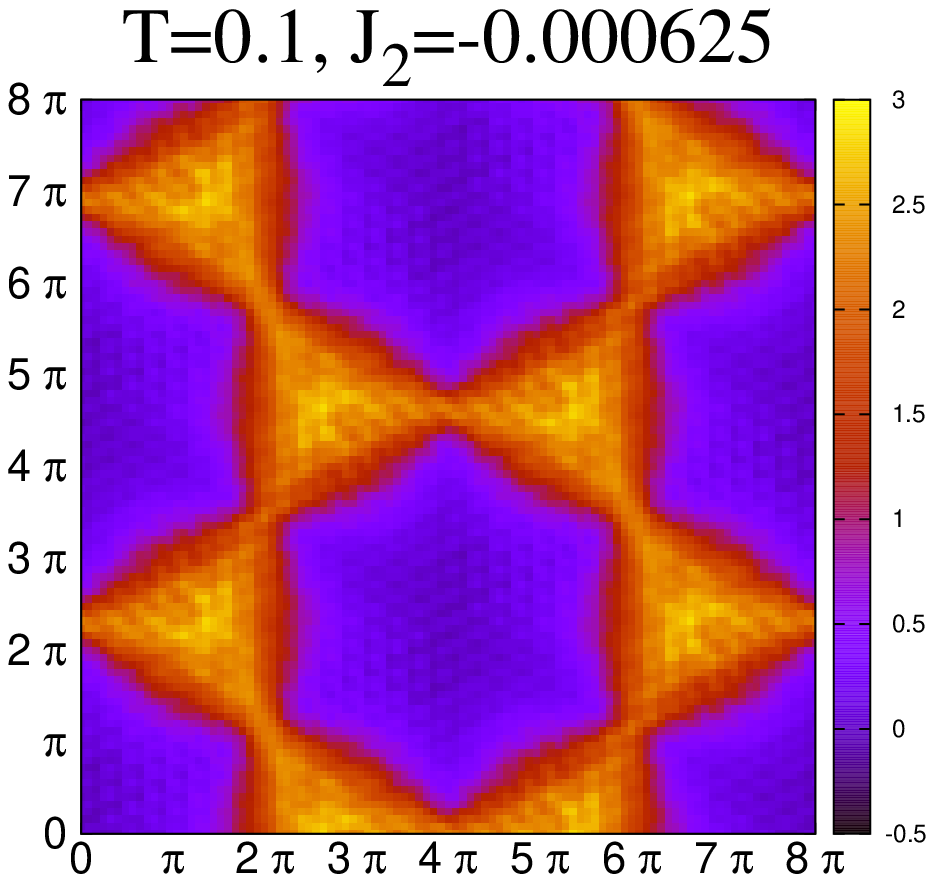}}  &
      \resizebox{50mm}{!}{\includegraphics{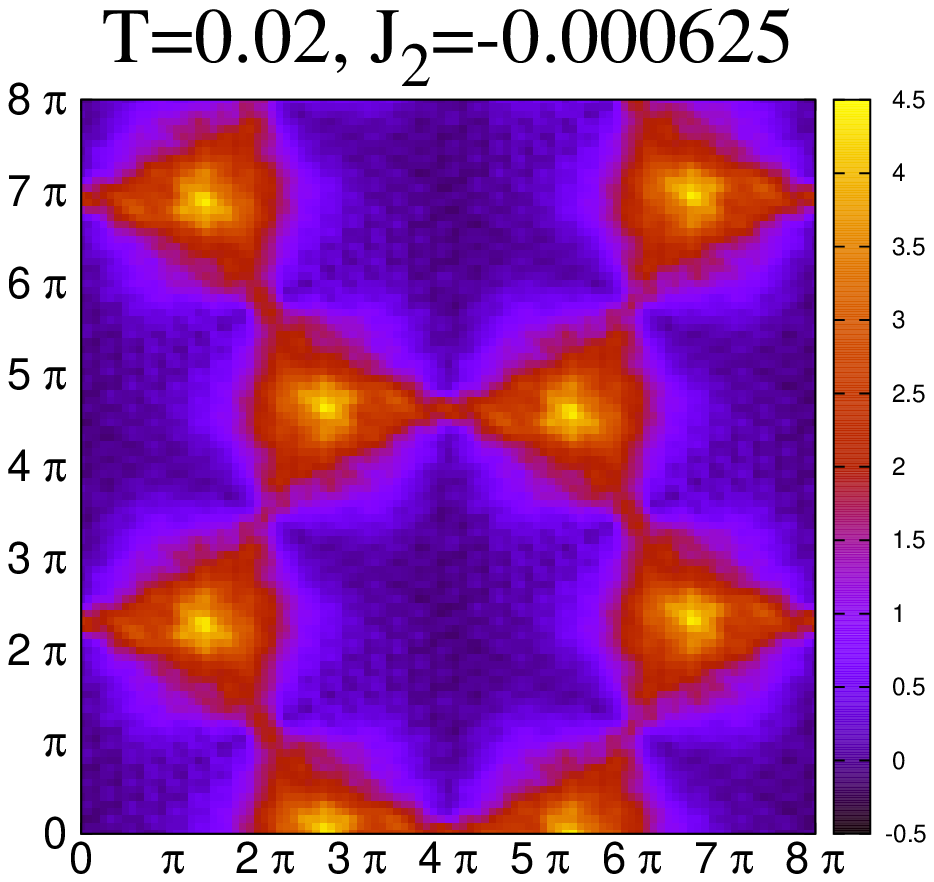}} &
      \resizebox{50mm}{!}{\includegraphics{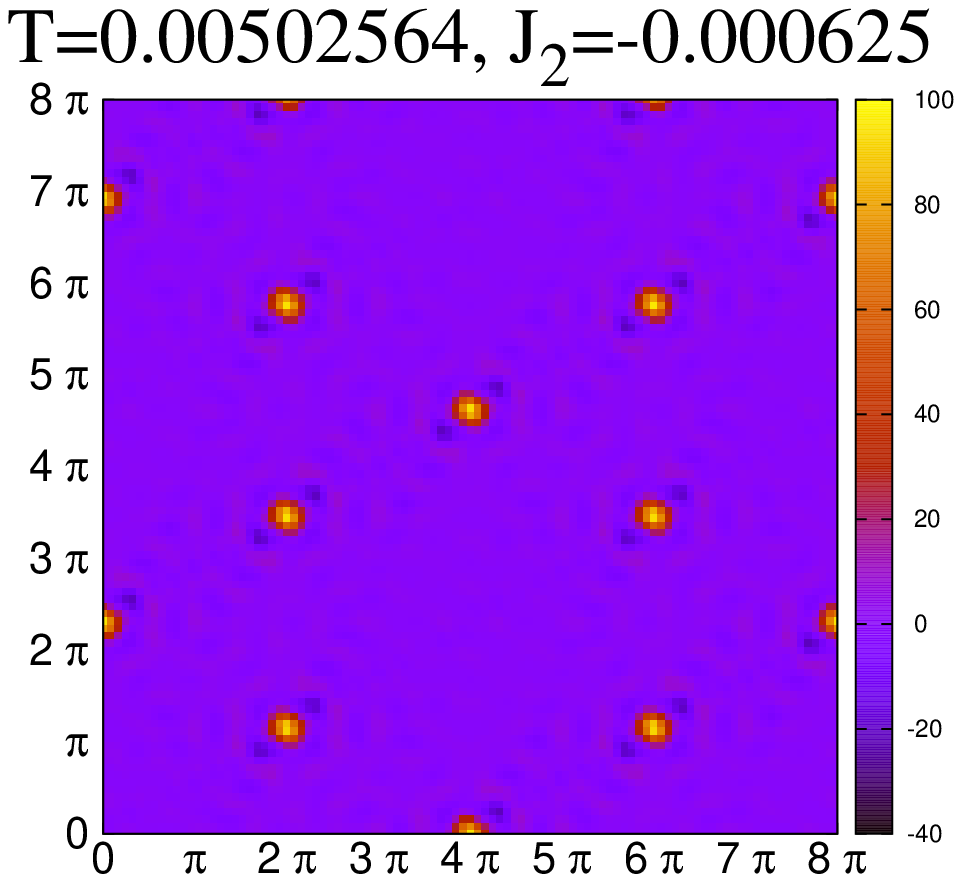}} \\
      \resizebox{50mm}{!}{\includegraphics{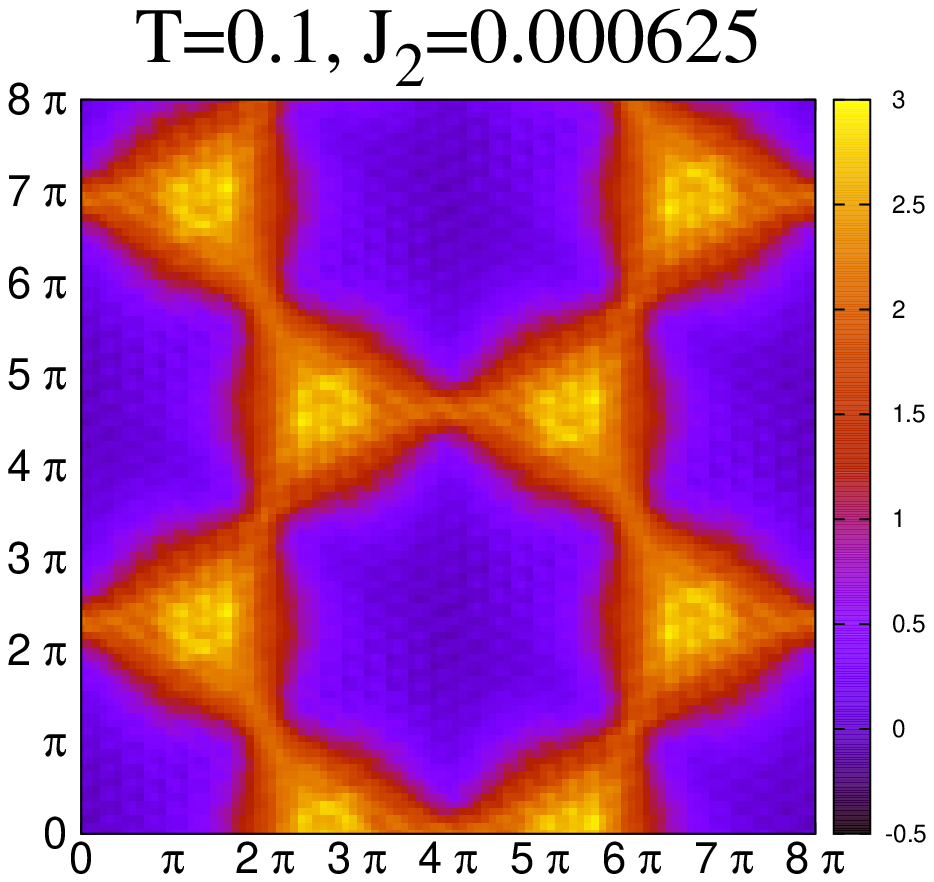}}  &
      \resizebox{50mm}{!}{\includegraphics{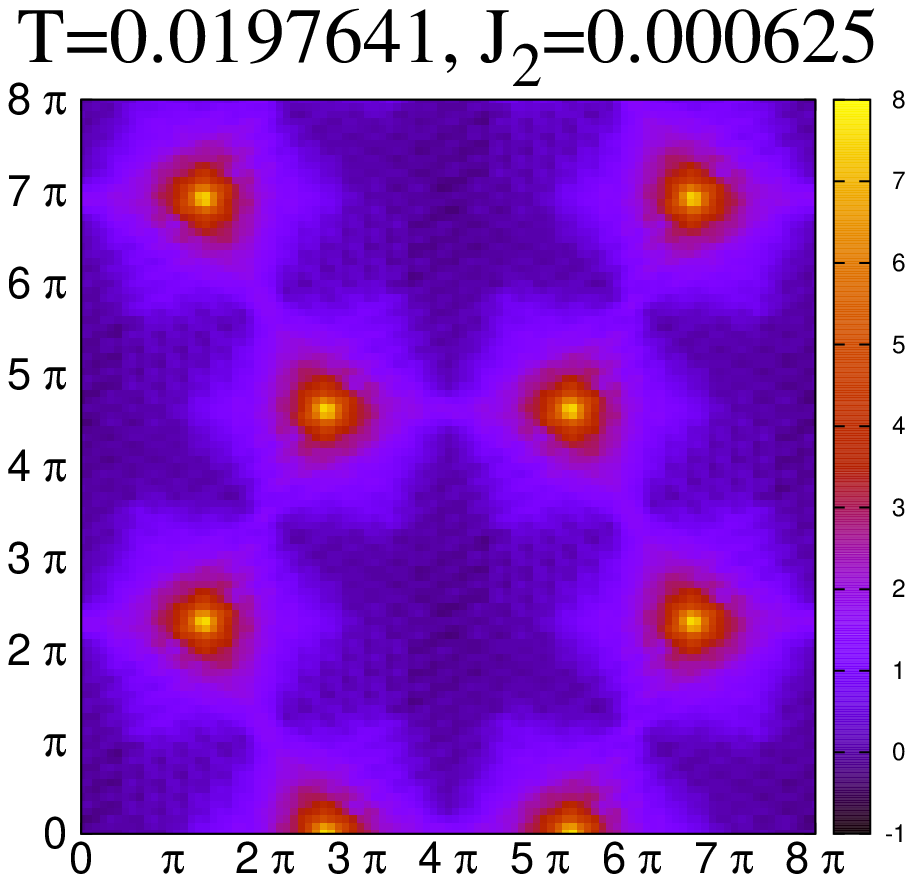}} &
      \resizebox{50mm}{!}{\includegraphics{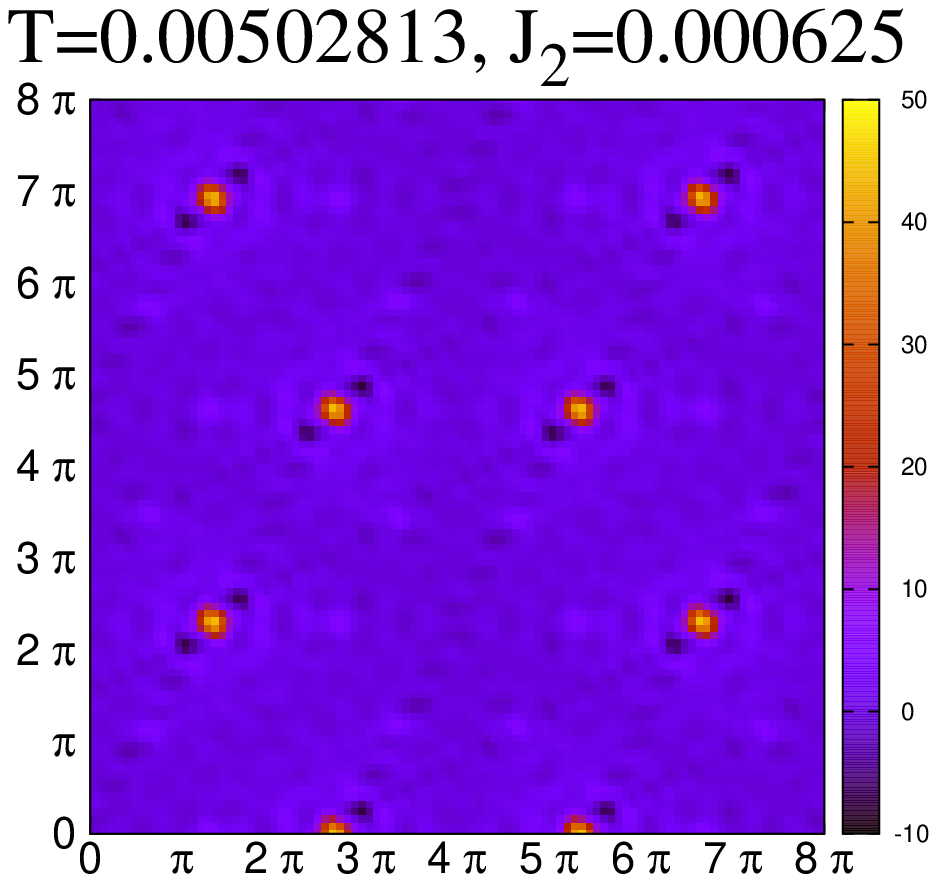}} \\
      \resizebox{50mm}{!}{\includegraphics{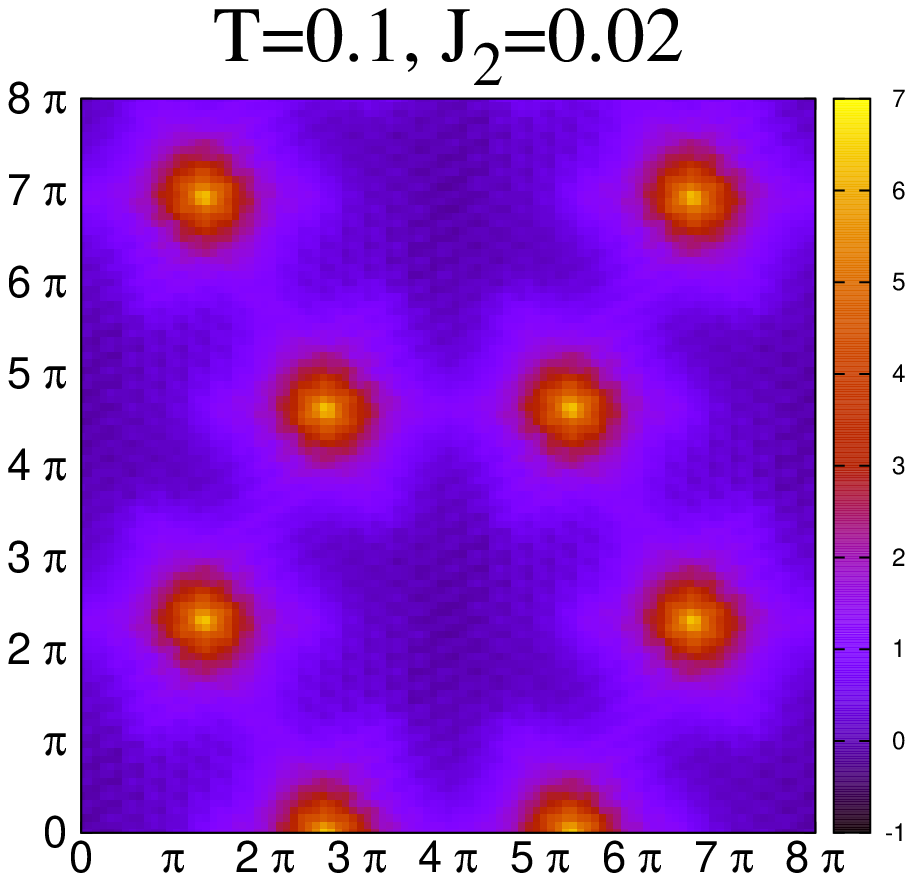}}  &
      \resizebox{50mm}{!}{\includegraphics{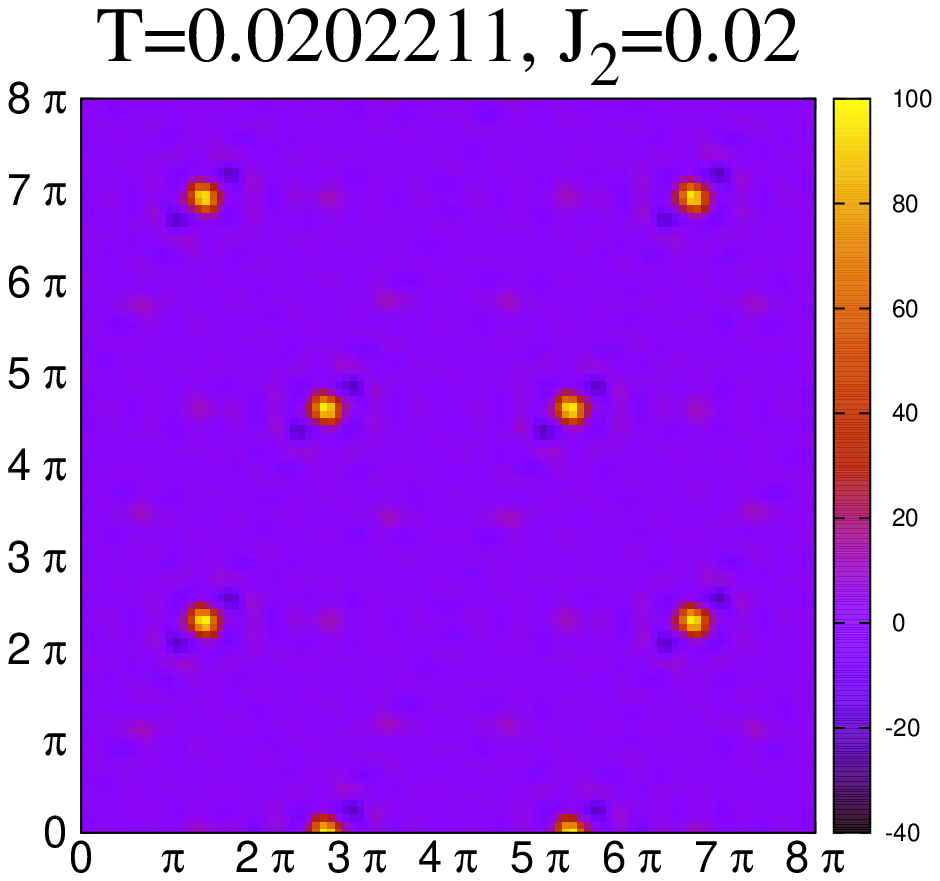}}  \\
    \end{tabular}
    \caption{Magnetic structure factor for $L=9$ systems for several values of $J_2$ and for the relevant temperature range.}
    \label{SF}
  \end{center}
\end{figure*}

We investigated the magnetic structure factor, to see the impact of the inclusion of the $J_2$-term on this quantity. For large $|J_2|$ we have chosen two temperatures, one above the transition ($T=0.1$) and one below ($T=0.02$). For the small $|J_2|$ cases we chose two more very low temperatures $T=0.005$ and $T=0.002$, the structure for these two very low temperatures is the same, therefore we included only $T=0.005$ in Fig. \ref{SF}. As the Kagome lattice is not bi-partite, the structure factor extents over several Brilouin zones. For high temperatures we observe the diffuse Bow-tie structure, as in the pure Kagome Lattice,\cite{ZHIT1} and as in related lattices such as the Pyrochlore.\cite{GAR} The structure appears for all investigated values for $J_2$ but at different temperatures. Lowering the temperature,
from there we find qualitatively the same structure than for the $J_2=0$ case for positive sign of $J_2$ (3 peaks at the center of the bow-tie prevailing). For negative $J_2$ with the other AFM ordering for the low-temperature phase, three peaks at the boundary of the triangle of the bow-tie develop, when lowering the temperature. In both cases the lowest energy plots finally show pinch points at the AFM ordering vectors for the state concerned but with an increased periodicity, while at high temperature the correlations are similar to the $J_2=0$ case. This can be seen from our plots in Fig. \ref{SF}: For e.g. the tiny values of $|J_2|$ shown ($J_2=-0.000625$ and $J_2=0.000625$) at $T=0.1$ we are above the transition and observe the Bow-tie structure, from there the pinch points develop when lowering this temperature to around $T=0,005$. For larger $|J_2|$ the temperature $T=0.1$ is already within the AFM phase, therefore one sees the development of the pinch points already at this temperature.
This means that aside of very low temperatures the additional $J_2$ term seem not to have much impact on the correlations.

\section{Conclusion} 

To sum up, we have extended earlier work on the classical Heisenberg model on the Kagome lattice, by studying the inclusion of next-nearest neighbor interactions. This is relevant for real systems and a good way to gain further insight to the nearest neighbor model. We see that tiny values of $J_2$ have a big impact on this model by enabling finite-temperature phase transition, with larger transition temperature compared to the crossover point for $J_2=0$. This suggests that in real materials having large spin on this lattice AFM order will be observed. The nature of the phase-transition has been explored, and it was found to fit the 2-D Ising universality-class for $|J_2|\ge0.01$, and shows consistency with a first order phase transition at the negative branch close to $J_2=0$. As the underlying induced symmetry is $Z_2$ an Ising transistion appears reasonable. How this transistion changes towards first-order might be an intersting point for future studies. For the tiny and positive $J_2$ case, two peaks in the specific heat develop, the second (in terms of higher temperature) most likely connecting further to the crossover point for the $J_2=0$ case. This is an interesting point for the nearest neighbor model and a possible line of attack for future studies. We notice that the lower transition line for the positive $J_2$ branch between the AFM 2 state and the disordered phase tend to $T\rightarrow0$ for $J_2\rightarrow0$, while in the other branch the AFM 1 order seems to vanish before reaching the $J_2=0$ point. This is consistent with earlier claims of a $T=0$ critical point for the AFM 2 transition. There seem to be no qualitative impact on the spin structure factor when tuning $J_2$. The region very close to $J_2=0$ remains still very interesting and should be further studied by  analytic and numeric methods.\\ 

{\it Acknowledgements: } We wish to thank S. Todo for advise on parapack and Parallel tempering. We are grateful for useful discussions with H. Monien and email correspondence with J. Chalker. The calculations have been carried out on the HPC facilities at the RWTH-Aachen.

\end{document}